\documentclass{PoS}

\title{Two-pion excited state contribution to pseudo-scalar correlators}

\ShortTitle{Two-pion excited state contribution}

\author{Oliver B\"ar\\
Institut f\"ur Physik,
\\Humboldt Universit\"at zu Berlin\\
12489 Berlin, Germany\\
E-mail: \email{obaer@physik.hu-berlin.de}}

\author{\speaker{Maarten Golterman}\\

        Institut de F\'\i sica d'Altes Energies, 
Universitat Aut\`onoma de Barcelona\\
E-08193 Bellaterra, Barcelona, Spain\\
{\rm and}\\
Department of Physics and Astronomy, San Francisco State University\\
San Francisco, CA 94132, USA\\
E-mail: \email{maarten@stars.sfsu.edu}}

\abstract{We study multi-particle state contributions to the QCD two-point functions of pseudo-scalar quark bilinears in a finite spatial volume. For sufficiently small quark masses  one expects three-meson states with two additional pions at rest to have the lowest total energy after the ground state.
Using chiral perturbation theory, we find the amplitude of this state to be too small to be seen
in present-day lattice simulations.  We speculate that curvature in the
effective mass plot extracted from the pseudo-scalar density two-point function instead corresponds to a genuine resonance, the $\pi(1300)$.}

\FullConference{The 7th International Workshop on Chiral Dynamics\\

                 August 6 -10, 2012\\

                 Jefferson Lab, Newport News, Virginia, USA}

\begin{document}

\section{Introduction}
In this talk, we review recent work on excited multi-particle states with
two additional pions in addition to the ground-state hadron \cite{BG}.

In lattice QCD, hadron masses are extracted by computing two-point
time correlation functions of local operators with the desired quantum numbers.
The two-point correlator for a meson operator $P$ in lattice QCD (we will
assume a vanishing spatial momentum) can be written as
\begin{equation}
\label{corr}
C(t)=\langle P(t)P^\dagger(0)\rangle=\sum_{n=0}^\infty C_n\,e^{-E_n t}\ .
\end{equation}
On the lattice, one works
in a finite spatial volume $L^3$, with $L$ the linear size of the lattice volume.   We will assume that the time extent of
the lattice is infinite, an assumption which is justified for most lattice QCD
computations, in which this time extent is much larger than $L$.

While the ground state is a stable particle (for instance, a pion or kaon), excited
states, labeled by values of $n>0$, can correspond to single-particle states
(resonances, such as the $\pi(1300)$), or multi-particle states (such as a state
with three pions).   In order to disentangle the spectroscopy of these excited
states, it will be necessary to tell these two different possibilities apart.

When the (up and down) quarks have sufficiently small masses, the first
excited state in the sum over states in Eq.~(\ref{corr}) will be a multi-particle
state composed of the ground state with two additional pions, with an energy
approximately equal to $E=E_0+2m_\pi$.

Consider, for example, the results found in Ref.~\cite{CERN1}, some of which
are shown in Fig.~1, taken from that paper.   This figure shows two
effective-mass plots of a valence meson in QCD with two dynamical flavors.
The valence-meson mass is almost the same in the left and right panels, but the sea-quark mass
is different, with the lighter sea quark in the right panel.   If only the ground-state meson mass would be visible, the effective mass would be a horizontal
line (indicated by the grey bands).  The fact that the lattice data do not fall on
this line indicates the presence of higher excited states, and it was conjectured in Ref.~\cite{CERN1} that the visible excited state is a multi-particle state with two extra pions in addition to the ground-state valence meson.  Indeed, a good fit was reportedly obtained with
\begin{equation}
\label{effmass}
M_{\rm eff}(t)=-\frac{d\ }{dt}\log{C(t)}=M_K\left(1+c\,e^{-(M'-M_K)t}\right)\ ,
\end{equation}
in which $M_K$ is the mass of the ground-state meson, $M'=M_K+2M_\pi$
with the pion $\pi$ made out of sea quarks, and $c$ a free constant.
The question we wish to address here is whether the
interpretation of the second exponential as due to a state with two extra
pions at rest is correct.

\begin{figure}[t]
\centering
\includegraphics[width=6in]{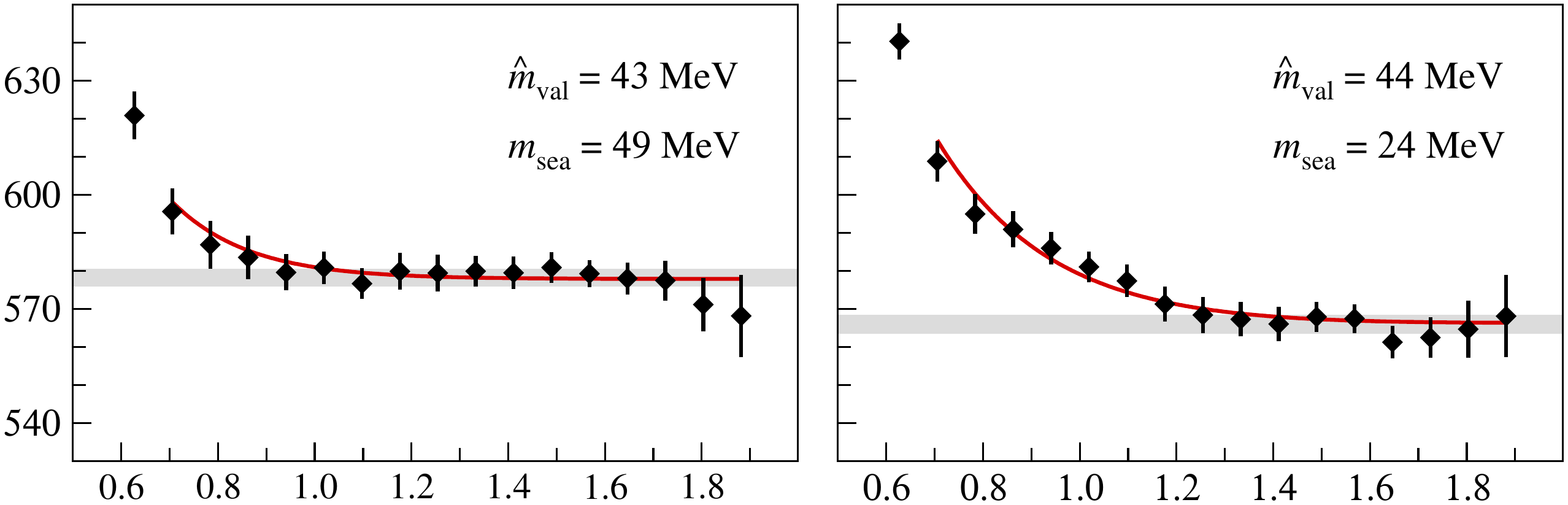}
\vspace*{-4ex}
\caption{Effective-mass plots of ``valence-meson'' mass, which is held (almost)
fixed, with two different sea-quark masses (in partially quenched QCD), from
Ref.~\cite{CERN1}.}
\vspace*{2ex}
\end{figure}

\section{Multi-particle states in a finite volume}
For three-particle state with two extra pions, one expects the contribution to the two-point
correlator~(\ref{corr}) to look like \cite{BG}
\begin{equation}
\frac{1}{L^6}\sum_{p,q,k}\delta_{p+q+k,0}\frac{1}{8E_p^\pi E^\pi_q E^K_k}
|\langle 0|P(0)|\pi(p)\pi(q)K(k)\rangle|^2\,e^{-E_{tot}t}\ ,
\end{equation}
where $p$, $q$ and $k$ are the spatial momenta of the three mesons,
and $E_{tot}\approx E_p^\pi+ E^\pi_q+ E^K_k$ as long as the momenta
are small, and the mesons weakly interacting.
The lowest state has $p=q=k=0$.   Moreover, in a finite volume, the state
with the smallest non-vanishing momentum has
\begin{equation}
E_p^\pi=E_q^\pi=M_\pi\sqrt{1+\left(\frac{2\pi}{M_\pi L}\right)^2}\ ,
\qquad p = -q = 2\pi/L\ ,
\end{equation}
and is thus typically much suppressed: for a typical value of $M_\pi L=4$,
the square-root factor is approximately equal to two.

Of course, in order to check whether a three-particle state such as 
hypothized in Ref.~\cite{CERN1} explains the effective mass seen in
Fig.~1, one could vary the volume, {\it i.e.}, $L$.   However, 
for light-enough pions, we can calculate the coefficient $c$ in Eq.~(\ref{effmass}) in chiral perturbation theory (ChPT), and check the
quantitative contribution of the three-particle states in more detail at
fixed $L$.
In Ref.~\cite{BG} we carried this out in lowest-order (LO) ChPT with
three light flavors.\footnote{Note that partial quenching is not relevant
when the sea-quark mass is smaller than the valence quark mass
\cite{BG}.}

For example, taking $P=\overline{d}\gamma_5 u$ in Eq.~(\ref{corr}), we find
\begin{equation}
\label{corrchpt}
C(t)=-\frac{f^2B^2}{2M_\pi}\,e^{-M_\pi t}\left(1+\frac{45}{512(fL)^4
(M_\pi L)^2}\,e^{-2M_\pi t}\right)\ ,
\end{equation}
where $f=f_\pi=f_K$ is the meson decay constant to LO (in the 
normalization in which $f_\pi\approx 92.2$~MeV), and $Bf^2$ is the
LO quark condensate.\footnote{In Ref.~\cite{BG} several other cases
were considered.}    Note that Eq.~(\ref{corrchpt}) contains the
leading three-particle state only, in which the two additional pions
are at rest.   There are, of course, other states with non-zero momentum
particles, additional kaon pairs, {\it etc}.   Our power counting is such that
we are in the $p$-regime, with
\begin{equation}
M_\pi\sim M_K\sim p\sim 1/L\ .
\end{equation}
With this power counting, the diagrams contributing to Eq.~(\ref{corrchpt})
are shown in Fig. 2; all other diagrams are of higher order.

\begin{figure}[tp]
\begin{center}
\includegraphics[scale=0.45]{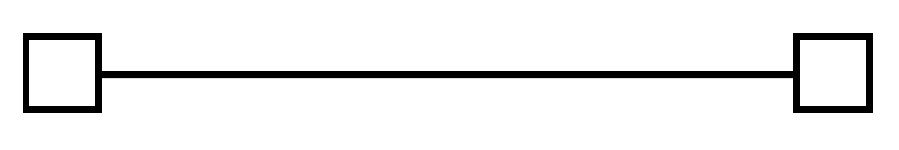}\hspace{0.5cm}$\null^+$\hspace{0.5cm}
\includegraphics[scale=0.45]{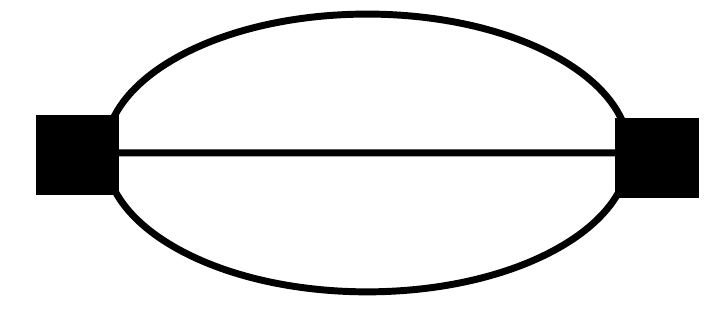}\hspace{0.5cm}$\null^+$\hspace{0.5cm}
\includegraphics[scale=0.45]{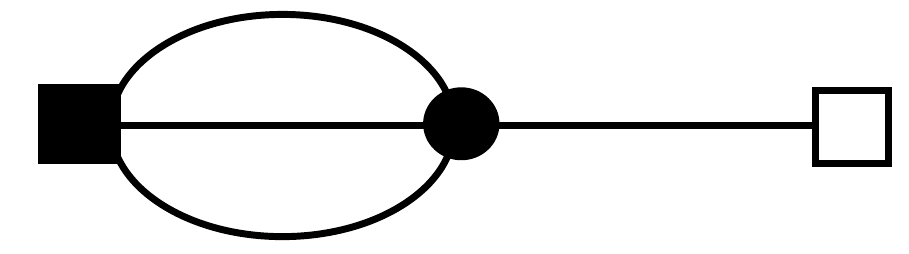}\\
\vspace{3ex}$\null^+$\hspace{0.5cm}
\includegraphics[scale=0.45]{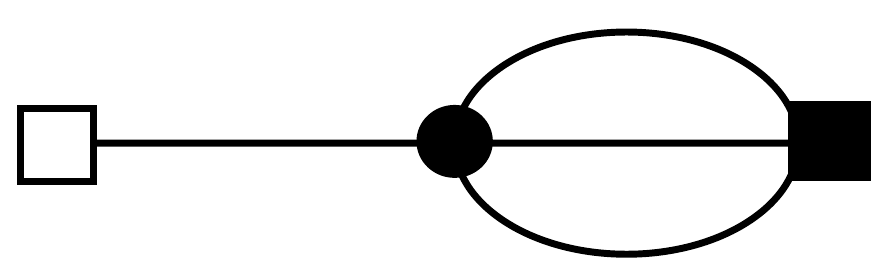}\hspace{0.5cm}$\null^+$
\hspace{0.5cm}\includegraphics[scale=0.45]{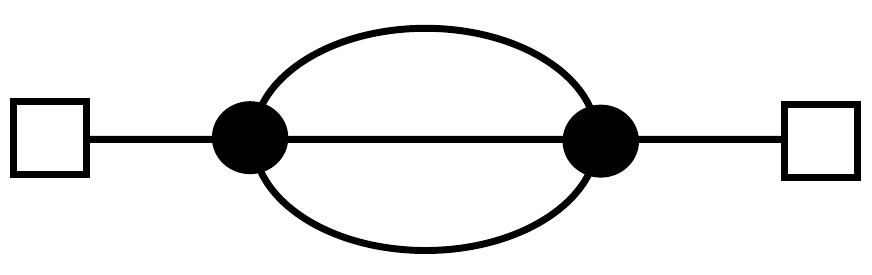}\\
\caption{Feynman diagrams for the $P$ correlation function. The squares represent the operators $P$ and $P^\dagger$ at times $t$ and $0$, where the open and solid squares denote the one-pion and three-pion terms, respectively. The circles represent a vertex insertion; an integration over these points is implicitly assumed. At least two of the three lines connecting a solid square or circle are pion propagators for the
contribution shown in Eq.~(2.3).}
\end{center}
\end{figure}

\section{Examples}
Now, let us return to Fig.~1, left panel, for which all meson masses are
approximately equal.   For this result, one finds from the details provided in Ref.~\cite{CERN2} that $fL\approx 1$, $M_\pi L\approx 5.9$.
For these values, $c\approx 5\times 10^{-3}$, which is much too small to
explain the curve, for which $c$ would have to be more than two orders
of magnitude larger.

In this example, $M_\pi\approx 620$~MeV, and one may worry that this is too
large for LO ChPT to give reliable results.   However, we find similar
conclusions for the right panel in Fig.~1, for which $M_\pi\approx 420$~MeV.

A very similar computation with a lighter pion mass ($M_\pi=270$~MeV)
was recently carried out in Ref.~\cite{ALPHA}.   We show the relevant 
results in Fig.~3.   In this case, $c=1.6\times 10^{-3}$, leading to the blue
curve shown in the figure.   Again, this value is much to small to explain the
data.

\begin{figure}[t]
\begin{center}
\includegraphics[scale=0.75]{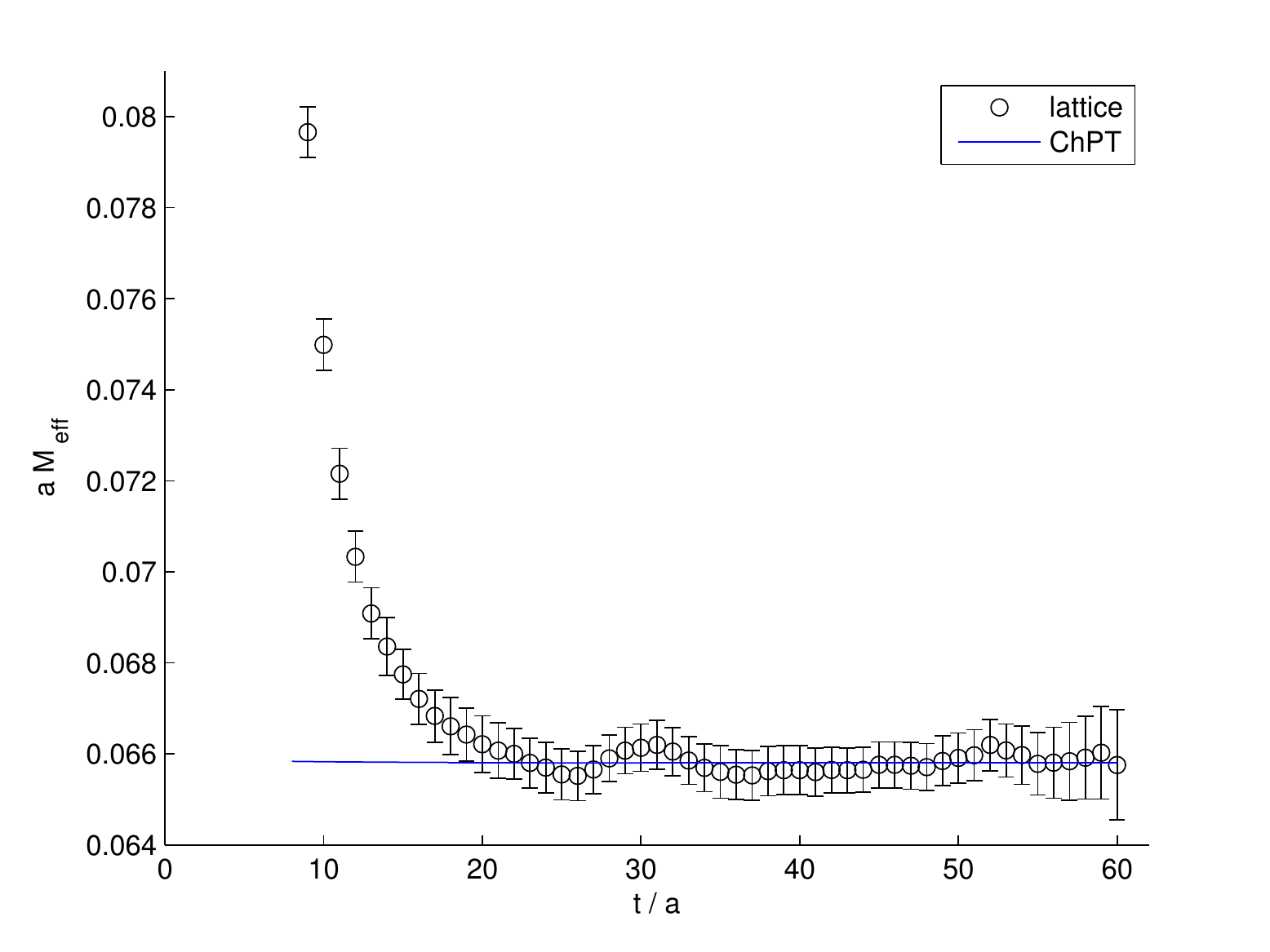}
\caption{The effective mass for the O7 data of Ref.~\cite{ALPHA} (also shown in Fig.~2 of this reference.) The solid blue line corresponds to the ChPT predicition for the three-pion state contribution to the effective mass.}
\end{center}
\end{figure}

\section{Speculation}
Of course, our conclusions above raise the question what excited states
could instead explain the data seen in Figs.~1 and 3.   Several possibilities come to mind, {\it a priori}.   The signal could be due to a two-particle state, like a
$\rho\pi$ or even a $\sigma(500)\pi$ state.   For such a state, the coefficient
$c$ in Eq.~(\ref{effmass}) would be suppressed only by a factor $1/L^3$
instead of a factor $1/L^6$.   Nevertheless, we think that these explanations
are unlikely.   {}From the extensive data reported in Ref.~\cite{CERN2},
we estimate that the two-particle energy of a $\rho\pi$ state (which has to be in a state with orbital angular momentum equal to one) would be
at least equal to $2$~GeV, too large to fit the time dependence of Fig.~1.
It is also likely that the overlap of a $\sigma(500)\pi$ state with the local
meson operators employed in Refs.~\cite{CERN1,CERN2} is too small.

An alternative explanation is then, of course, that the curvature seen in
Fig.~1 is due to a genuine resonance.   If we parametrize the effective mass
by 
\begin{equation}
M_{\rm eff}=M\left(1+A\,e^{-(M'-M)t}\right)\ ,
\end{equation}
we find, by eye, from the red curves in Fig.~1 that
\begin{eqnarray}
\label{MprimeA}
M'&=&1.67\ \mbox{GeV}\ ,\qquad A=1.8\qquad \mbox{(left\ panel)}\ ,\\
M'&=&1.45\ \mbox{GeV}\ ,\qquad A=2.2\qquad \mbox{(right\ panel)}\ .
\nonumber
\end{eqnarray}
Of course, both $M'$ and $A$ are dependent on the sea pion mass.
To LO in ChPT, this dependence would be likely be linear in $M_\pi^2$:
\begin{equation}
M'(M_\pi^2)=M'_0+b M_\pi^2\ ,\qquad A(M_\pi^2)=A_0+a M_\pi^2\ .
\end{equation}
{}From the values in Eq.~(\ref{MprimeA}), we then find
\begin{equation}
M'(M_\pi=140\ \mbox{MeV})=1.27\ \mbox{GeV}\ ,\qquad
b=1.1\ \mbox{GeV}^{-1}\ ,\qquad a=-1.9\ \mbox{GeV}^{-2}\ .
\end{equation}
The value for $M'$ at $M_\pi=140$~MeV is quite close to the mass
of the $\pi(1300)$, and the values of $a$ and $b$ are of order the 
typical hadronic scale.   It might thus well be the case that the
curvature seen in Fig.~1 is due to a genuine physical resonance. 

\section{Conclusions}
The spectroscopy of hadronic excited states is complicated, and it is 
important to use all available tools in order to further the recent progress
\cite{reviews}.   In order to identify the lowest multi-particle states in a 
given channel, ChPT can be helpful, because in many cases,
for small enough quark
masses, those multi-particle states are composed of the ground-state
hadron in that channel plus two or more pions.   Since hadronic states
with different numbers of pions are related by chiral symmetry, the 
contributions of such states can be calculated in ChPT.   It follows that
the examples discussed in this talk should generalize to other channels,
such as for instance the nucleon.

In a finite spatial volume, the state with two extra pions is suppressed by the
square of the spatial volume.   For pseudo-Goldstone boson channels, 
this suppression is of order $(f_\pi L)^{-4}(M_\pi L)^{-2}$ times a numerically
small factor of order $1/10$ (if axial currents instead of pseudo-scalar
densities are used, the suppression is even stronger).   We find, therefore,
that such contributions can be ignored in the analysis of currently typical
lattice data for hadron spectroscopy, when local operators are employed.

\vspace{3ex}
\noindent {\bf Acknowledgments}
\vspace{3ex}

We thank Martin L\"uscher, Stefan Schaefer, Francesco Virotta for correspondence and Rainer Sommer and Hank Thacker for discussions. OB is supported in part by the Deutsche Forschungsgemeinschaft (SFB/TR 09). MG is supported in part by the US Department of Energy
and by the Spanish Ministerio de Educaci\'on, Cultura y Deporte, under program SAB2011-0074.
MG also thanks the Galileo Galilei Institute for Theoretical Physics for
hospitality, and the INFN for partial support.

\end{document}